\documentclass[%
reprint,
showpacs,preprintnumbers,
amsmath,amssymb,
aps,
prl,
]{revtex4-1}

\usepackage{graphicx}
\usepackage{dcolumn}
\usepackage{bm}

\usepackage{color}

\begin{document}

\preprint{APS/123-QED}

\title{Generation of high-frequency strain waves during femtosecond demagnetization of Fe/MgO films}

\author{T Henighan$^{1,2}$}\thanks{henighan@slac.stanford.edu}
\author{M Trigo$^{1,3}$}
\author{S Bonetti$^3$}
\author{P Granitzka$^{3,4}$}
\author{D Higley$^{3,5,6}$}
\author{Z Chen$^{2,3}$}
\author{M P Jiang$^{1,2}$}
\author{R Kukreja$^3$}
\author{A Gray$^3$}
\author{A H Reid$^3$}
\author{E Jal$^3$}
\author{M C Hoffmann$^6$}
\author{M Kozina$^{1,6}$}
\author{S Song$^6$}
\author{M Chollet$^6$}
\author{D Zhu$^6$}
\author{P F Xu$^{7,8}$}
\author{J Jeong$^7$}
\author{K Carva$^{9,10}$}
\author{P Maldonado$^{10}$}
\author{P M Oppeneer$^{10}$}
\author{M G Samant$^7$}
\author{S S P. Parkin$^{7,8}$}
\author{D A Reis$^{1,3,5}$}
\author{H A D{\"u}rr$^3$}\thanks{hdurr@slac.stanford.edu}

\affiliation{$^1$PULSE Institute, SLAC National Accelerator Laboratory, Menlo Park, California, USA}

\affiliation{$^2$Physics Department, Stanford University, Stanford, California, USA}


\affiliation{$^3$Stanford Institute for Materials and Energy Sciences, SLAC National Accelerator Laboratory, 2575 Sand Hill Road, Menlo Park, CA 94025.} 

\affiliation{$^4$Van der Waals-Zeeman Institute, University of Amsterdam, 1018XE Amsterdam, The Netherlands}

\affiliation{$^5$Department of Photon Science and Applied Physics, Stanford University, Stanford, California, USA}

\affiliation{$^6$Linac Coherent Light Source, SLAC National Accelerator Laboratory, Menlo Park, California, USA}

\affiliation{$^7$IBM Almaden Research Center, 650 Harry Road, San Jose, California 95120, USA}

\affiliation{$^8$Max-Planck Institute for Microstructure Physics, 06120 Halle (Saale), Germany}

\affiliation{$^9$Charles University, Faculty of Mathematics and Physics, Department of Condensed Matter Physics, Ke Karlovu 5, CZ-12116 Prague 2, Czech Republic}

\affiliation{$^{10}$Department of Physics and Astronomy, Uppsala University, P. O. Box 516, S-75120 Uppsala, Sweden}

\date{\today}

\begin{abstract}
We use femtosecond time-resolved hard x-ray scattering to detect coherent acoustic phonons excited during ultrafast laser demagnetization of bcc Fe films. We determine the lattice strain propagating through the film through analysis of the oscillations in the x-ray scattering signal as a function of momentum transfer. The width of the strain wavefront is $\sim$100 fs, similar to demagnetization timescales. First-principles calculations show that the high-frequency Fourier components of the strain, which give rise to the sharp wavefront, could in part originate from non-thermal dynamics of the lattice not considered in the two-temperature model.
\end{abstract}


\maketitle

The speed limits for collective spin, electronic and lattice motions are of fundamental interest and could have a profound effect on the ability to store and process information. So far the fastest manipulation of magnetic moments in ferromagnetic films has been achieved using femtosecond optical laser pulses \cite{Beaurepaire,Stanciu,Radu,Lambert}. Ultrafast demagnetization on timescales of only several hundred femtoseconds \cite{Beaurepaire,Koopmans} is an important ingredient in all-optical magnetic switching \cite{Stanciu,Radu}. Intriguingly magnetic switching using strong magnetic and electric field pulses takes place on similar timescales to ultrafast demagnetization \cite{Back,Gamble}. However, the underlying non-adiabatic motion of electrons and spins far from equilibrium and especially their coupling to the initially unperturbed lattice still poses a significant challenge to theory \cite{Koopmans,Schellekens,Carva}. Typically electron-phonon energy transfer following femtosecond laser heating in metals is described using the two-temperature model (2TM) \cite{Wright} or when including the spin system in a three-temperature model \cite{Beaurepaire,Koopmans}. These models have been used to explain ultrafast optical generation of lattice strain waves (coherent acoustic phonons) \cite{Wright} which can manipulate \cite{Kim} and coherently control \cite{Kim2} the magnetization orientation in ferromagnetic Ni films. Yet, the applicability of these models on short timescales remains to be proven. 

Femtosecond x-ray and electron scattering can provide a direct means for measuring the atomic-scale displacements associated with the propagating strain \cite{Reis2007,CaoPRL}. Nonetheless, experiments in metals have been limited primarily to observing the evolution of lattice temperature through the Debye-Waller factor \cite{CaoAPL} and the average lattice expansion through changes in the Bragg condition \cite{CaoPRL}. Although important for magneto-acoustic spin manipulation \cite{Kim, Kim2}, laser-induced strain waves have so far not been directly probed in magnetic 3d transition metals, and it is generally believed that the lattice strain is dominated by low-frequency waves leading to several ps long strain profiles \cite{Wright,Kim,Kim2}. 

Here we show that the strain wave duration can be significantly shorter than originally believed \cite{Kim}. We use femtosecond hard x-ray pulses to probe the temporal evolution of quasi-elastic Bragg scattering from coherent acoustic phonons to directly detect the frequency content of ultrafast lattice strain waves generated during the femtosecond laser demagnetization of ferromagnetic Fe/MgO(001) films. The observed coherent oscillations can be unambiguously assigned to a coherent acoustic phonon wavepacket with frequencies extending to 3.5 THz. The temporal width of the acoustic pulse is comparable to the observed electron and spin thermalization timescales \cite{Carpene2008,Carva}. Poor agreement is found when comparing the results to those of a 2TM which includes stress only from the heated lattice, particularly for high-frequency Fourier components of the strain. The inclusion of the stress originating from the electronic system in the 2TM dramatically improves the fit by providing sharper spatial features to the strain. However, the validity of the 2TM, which assumes a thermal distribution for the phonons, is questionable considering the phonon thermalization timescales of $\sim$10 ps are longer than the timescales probed here. While previous works have noted that the electron distribution is likely non-thermal during the first $\sim$100 fs \cite{Carpene2006}, the nonthermal behavior of the lattice has been largely ignored. Here we show \textit{ab initio} calculations suggesting the phonons are highly nonthermal for tens of picoseconds in iron. We also present \textit{ab initio} calculations of the nonequilibrium electron-phonon energy transfer which show that the stress generated by the non-thermal lattice can provide high-frequency Fourier components to the strain, similar to the electronic stress in the 2TM. These nonthermal dynamics may be important to consider in the ultrafast demagnetization of iron and other situations where the 2TM is employed.

The Fe layer was deposited on a MgO substrate and capped with a 3 nm layer of MgO to prevent oxidation. Further details on sample fabrication are provided in the supplement. Time-resolved magneto-optic kerr experiments established identical demagnetization behavior as observed previously \cite{Carpene2008}. The amount of demagnetization was less than 10\% for the pump fluence of about 1mJ/cm$^2$ used here. Optical pump – x-ray probe measurements were performed at the XPP instrument \cite{xppref} of the Linac Coherent Light Source free-electron laser with pink beam at 120 Hz repetition rate and $\sim$10$^{12}$ photons per pulse. The photon energy was set to 7 keV, just below the iron K edge to avoid fluorescence background. The x-ray scattering intensity was measured with an area detector \cite{cspad}. Optical 800 nm pump pulses were 60 fs in duration. The time delay between the optical pump and x-ray probe was corrected for the x-ray arrival time jitter on a shot-by-shot basis \cite{timingtool}. A custom quadrupole electromagnet was used to control the film’s in-plane magnetization direction. However, we observed no dependence of the diffraction data on the in-plane magnetization direction. We operated in a reflection geometry with an x-ray (optical) cross section of 11$\times$130 $\mu$m (300$\times$390 $\mu$m) projected onto the sample at a grazing angle of 0.4 (2.4) degrees to match the x-ray penetration depth and film thickness. The finite optical-x-ray crossing angle results in a negligible temporal smearing compared to the $\sim$100 fs resolution due to the finite durations of the pump and probe pulses. The optical pulses were p-polarized with respect to the sample to minimize reflection losses. Sample motion was restricted to rotations about the sample normal to preserve the grazing x-ray incidence angle. The x-ray scattering was measured along the conventional (01L) Bragg rod at different positions of L = 1 + $\rm q_z$. In the kinematic limit, diffraction from ultrathin films with N atomic planes consists of discrete satellites spaced $\sim$1/N in reciprocal lattice units from the main peak \cite{Warren}. We did not detect the individual satellite peaks for the 23 nm thick film used in this study because the x-ray spot size was kept large to avoid damage by the x-ray laser. However a similarly prepared 12 nm film displayed clearly separated satellite features in reference measurements at the Stanford Synchrotron Radiation Lightsource (not shown) attesting the excellent epitaxial quality of the Fe/MgO samples. 

\begin{figure}
\centering
\includegraphics[width=3.2in]{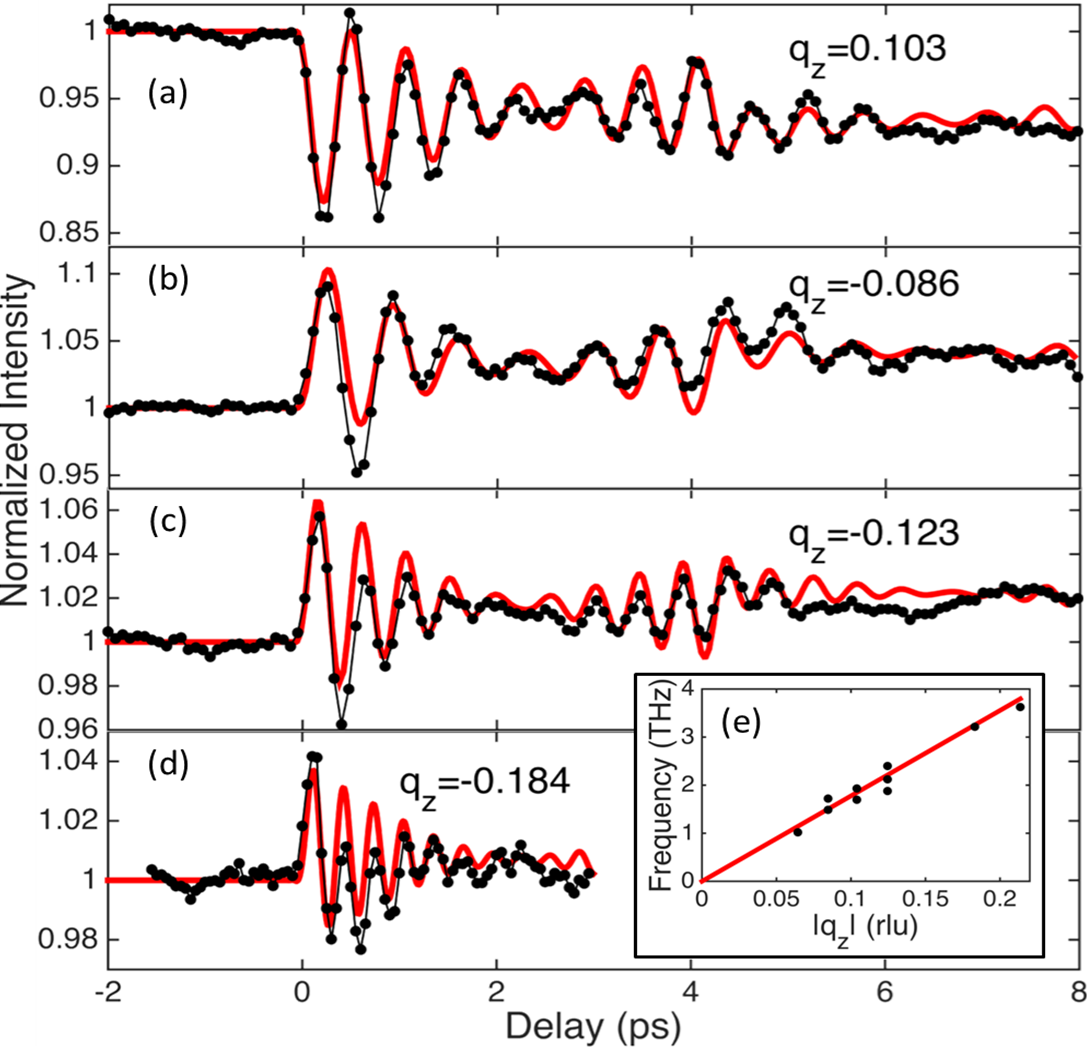}
\caption{(color online). Diffraction data (light lines and solid symbols) from a 23 nm thick Fe film as a function of optical pump- x-ray probe time delay for reduced wavevectors $\rm q_z$ (a-d). Heavy lines correspond to the best fit to the data for the laser-induced strain model described in the text. (e) shows the frequency of the x-ray intensity oscillations vs $\rm q_z$ (black dots) and the 5.13 nm/ps bulk speed of sound (red line).}
\label{fig:f1}
\end{figure}

Fig. \ref{fig:f1} (a-d) shows time-resolved diffraction traces (black lines and symbols) measured at different momenta transfer along the Bragg truncation rod (011+$\rm q_z$). The scattered intensity is integrated over a small region of reciprocal space encompassing 1-3 satellite peaks that are selected by the scattering geometry. Each trace shows high-frequency oscillations (up to 3.5 THz as shown in the inset) accompanied by a more slowly oscillating envelope. This beating is a result of the integration over multiple satellites, each with slightly different frequency. Fig. \ref{fig:f1} (e) displays the dominant frequencies as a function of $\rm q_z$. We find a linear relationship with the slope closely matching the bulk longitudinal speed of sound of 5.13 nm/ps along $<$001$>$ (red line) \cite{crc}.  This clearly indicates that the temporal oscillations are related to laser-excited longitudinal acoustic phonons traveling through the crystal along the film normal with wavevectors $\rm q_z$. Fig. \ref{fig:f1} also shows that the phonons initially oscillate in phase as expected for a coherent acoustic strain pulse generated by a stress that is nearly instantaneous when compared to a half-period of the highest frequency modes ($\lesssim$150 fs). The step just after 4 ps in figures 1(a-c) corresponds to the acoustic propagation time across the thickness of the film. At this time a portion of the strain wavefront originating at the free surface transmits into the substrate at the same time that the strain wavefront originating at the substrate reaches the cap-layer Fe/MgO interface.

\begin{figure}
\centering
\includegraphics[width=3.2in]{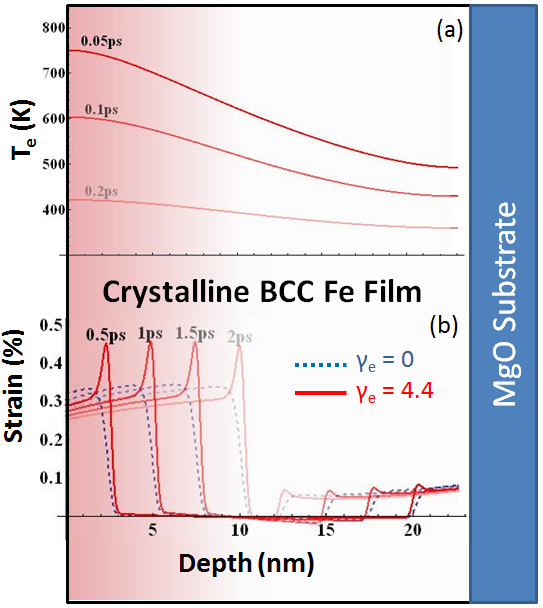}
\caption{(color online). Spatial profile of the electron temperature (a) and of the strain (b). The red solid lines (blue dashed lines) represent strain profiles obtained by fitting the experimental data to the model including (excluding) the effect of electronic stress, $\sigma_e$, from the laser-heated electronic system. 
}
\label{fig:f2}
\end{figure}

We next describe the modeling of the strain wave and its effect on the diffraction pattern. A 2TM described previously \cite{Wright} was used to simulate the temporal evolution of electron and lattice temperatures. Both quantities also show spatial variations due to the finite optical penetration depth as illustrated in Fig. \ref{fig:f2} (a). Finite carrier diffusion serves to diminish these spatial variations in time.  The result is a uniaxial but spatially and temporally varying stress, $\sigma(t,z)$, with contributions from not only the lattice, but also the electronic subsystem \cite{Thomsen}. The total stress is
\begin{multline}
\sigma(t,z) = \sigma_e(t,z) + \sigma_l(t,z) =\\-\int_{T_e(t=0)}^{T_e(t,z)}\gamma_e C_e(T_e') dT_e' - \int_{T_l(t=0)}^{T_l(t,z)} \gamma_l C_l(T_l') dT_l'
\label{eq:stress}
\end{multline}
where $z$ describes the distance from the cap-layer Fe/MgO interface, $C_{e,l}$ are the heat capacities and $\gamma_{e,l}$ the Gr{\"u}neisen parameters and $T_{e,l} (z,t)$ are the temperatures \cite{Wright, Thomsen}. The subscripts $e$ and $l$ refer to the electron and lattice subsystems, respectively.  In equilibrium, $\gamma_l$ and $\gamma_e$ are typically of the same order of magnitude, while $C_l$ exceeds $C_e$ by 2 orders of magnitude at room temperature \cite{Beaurepaire,Kim}, such that a fixed energy deposition into either the electronic or lattice subsystems would result in stress of similar magnitude. However, all the energy initially goes into the electronic system resulting in a large overshoot in $T_e$ and $\sigma_e (t,z)$ during the first picosecond following laser excitation. The resulting strain along the film normal is found by solving the one-dimensional elasticity equations which take the form of the wave equation for the atomic displacements with the stress gradient as a driving force \cite{Thomsen}.  The interface between the Fe and MgO cap layer (top) is approximated as free, while the transmission and reflection at the Fe/MgO substrate (bottom) is treated, as due to the acoustic impendence mismatch, in a continuum model.

We find that introducing the electronic stress in addition to the stress from the lattice dramatically improves the agreement between data and simulation \cite{Nicoul}. The calculated strain profiles are shown in Fig. \ref{fig:f2} (b) for $\gamma_e=0$ and 4.4 (corresponding to neglecting or including the electronic stress). In either case, strain waves are launched at both Fe/MgO interfaces. The top Fe/MgO interface leads to larger strain due to the higher temperatures there compared to the bottom Fe/MgO interface (see the calculated temperature profiles in Fig. \ref{fig:f2} (a)). However, the shape of the acoustic pulse, particularly the sharpness of the front is strongly affected by the electronic contribution to the stress.  In this case the overshoot in the electronic temperature effectively drives higher-frequency, shorter-wavelength vibrational modes and modifies the frequency spectrum of the ensuing strain pulse via $\sigma_e(t,z)$ in Eq. (\ref{eq:stress}), particularly at the highest frequencies. When we include the electronic stress, the resultant strain has sharper spatial features, corresponding to increased amplitudes of short-wavelength, high-frequency Fourier components. 

Diffraction patterns from the transiently strained film were simulated using a kinematic-diffraction model which included the effects of heating (Debye-Waller factor) on the Bragg peak intensity and the finite attenuation length of the x-rays [suppl]. Since all data were taken away from the Bragg condition, dynamical scattering effects could be neglected \cite{Batterman}. The calculated diffracted intensity was fit to the data for the four different scattering conditions shown in Fig. \ref{fig:f1}. The only material parameter extracted from the model was $\gamma_e$. The other free parameters in the fit (absorbed fluence, x-ray grazing angle, film thickness, laser arrival time, and sample orientation) were allowed to vary within uncertainties of the measurement.

The best fits of calculated scattering from strain profiles and experimental data (red lines in Fig. \ref{fig:f1}) yields a $\gamma_e$ of 4.4 (reduced $\chi^2$ of 9.08) with the electron-phonon coupling constant, $G$, held fixed at $5.5 \times 10^{18}$ $\rm W m^{-3} K^{-1}$ \cite{Lin2008,e-pconstweb2}. The errors in the scattering yield were estimated from the standard deviation of the measured scattering in Fig. \ref{fig:f1} before arrival of the laser pulse (negative time delays). To achieve a similar goodness of fit with $\gamma_e$ fixed at zero required increasing the electron-phonon coupling constant $G$ by three orders of magnitude. This is physically implausible and illustrates the necessity of a stress resulting from the electronic temperature or some other process with a similarly rapid timescale to explain the results. Although $\gamma_e = 4.4$ is about twice the equilibrium value \cite{egruneisen1980,egruneisen1965}, we can achieve a similarly good fit $\gamma_e = 2.5$ if $G=1 \times 10^{18}$ $\rm W m^{-3} K^{-1}$. Parameters $G$ and $\gamma_e$ are strongly coupled in the fitting process, making it difficult to assign meaningful error bars. Any difference from the equilibrium values is unlikely caused by the spin system since the observed demagnetization is less than 10\% and a corresponding increase of the spin temperature should be small \cite{Carpene2008}.

While the 2TM including the electronic stress captures the main features in the data, the reduced $\chi^2$ of 9 suggests that the model is incomplete. The assumption of a thermal electron system for early times ($\sim$100 fs) is likely invalid \cite{Carpene2006} and previous works have speculated how this might affect the strain \cite{Tzianaki}. However, typical phonon lifetimes ($\sim$10 ps) would suggest that the lattice takes much longer to thermalize than the electrons. To this end we have performed \textit{ab initio} calculations of the phonon spectrum and phonon lifetime, as well as of the non-equilibrium electron-phonon energy transfer, the resulting  $\partial \sigma / \partial t$ and the strain. We show in an illustrative case that the nonthermal phonon system can also provide sharp spatial features to the strain.

\begin{figure}
\centering
\includegraphics[width=3.2in]{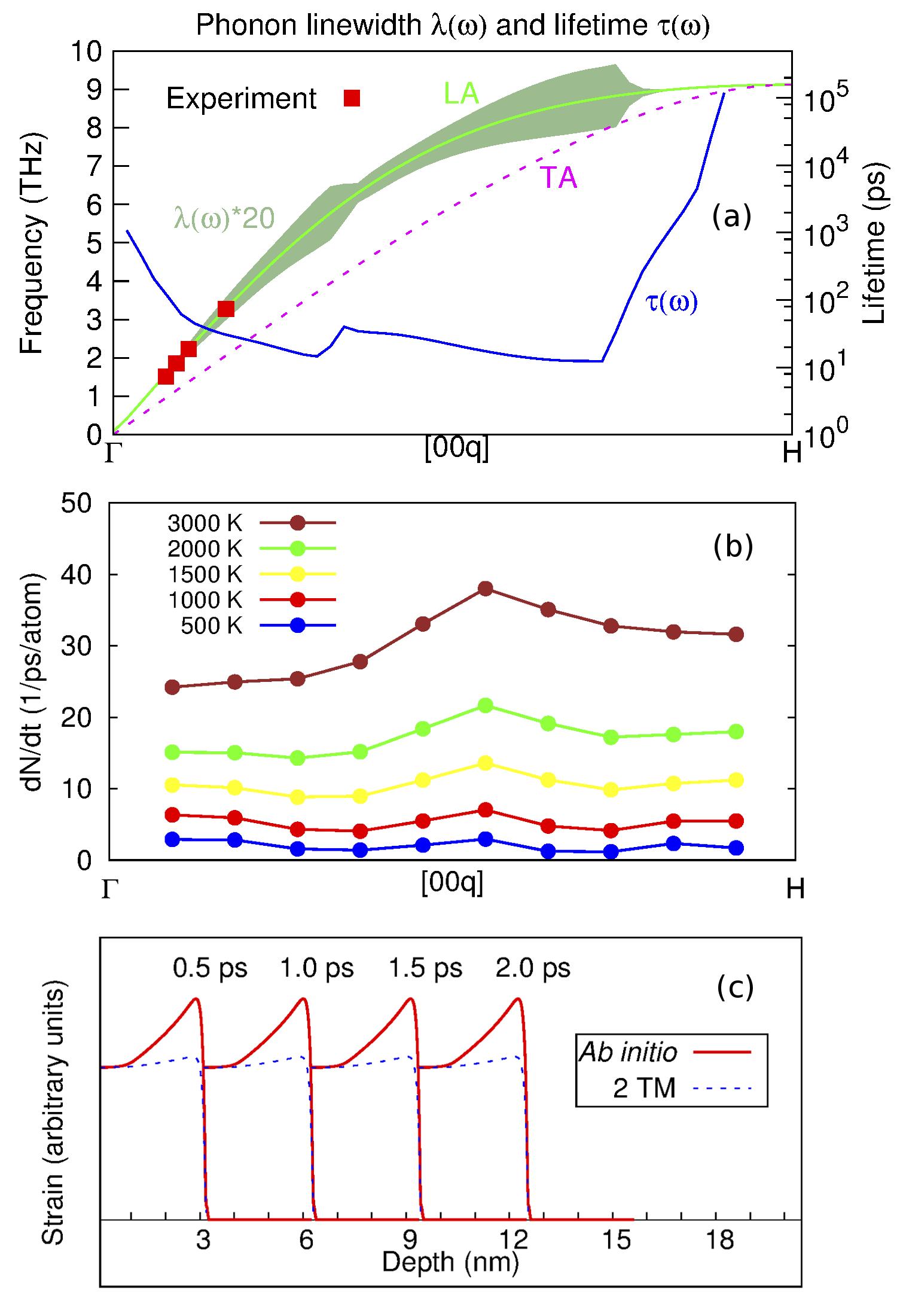}
\caption{(color online). (a) Calculated longitudinal (LA, green line) and transverse (TA, purple dashed line) acoustic phonon dispersions in bcc Fe along the $\Gamma-$H direction, including the LA inelastic phonon linewidth (green shaded area) and corresponding lifetime $\tau (\omega )$ (blue line). Red squares denote experimental values (from Fig.\ \ref{fig:f1}). (b) \textit{Ab initio} computed dependence of change of the phonon population $dN/dt$ on the electron temperature $T_e$ and $\bm{k}$-vector.
(c) The strain profiles computed for the indicated delay times from the \textit{ab initio} results in (b) and from the 2TM, both for $\gamma_e =0$. 
}
\label{fig:f3}
\end{figure}

Figure \ref{fig:f3}(a) shows calculated longitudinal acoustic (LA) and transverse acoustic (TA) phonon branches $\omega_{n\bm{k}}$ of Fe along the $\Gamma-$H direction. The lifetimes $\tau(\omega)$ of the LA phonons are larger than 10 ps, indicating that the phonon system remains out-of-equilibrium for all times reported in this Letter. Hence, in contrast to the 2TM, it cannot be assumed that a thermalized phonon system exists. Abandoning this approximation we write $\frac{\partial \sigma_l}{\partial t} = -\sum_{\bm{k}} \gamma_l ({\bm k}) \hbar \omega_{n\bm{k}} \frac{d N(T_e, \bm{k})}{dt}$, where $\gamma_l ({\bm k})$ are the mode-dependent Gr{\"u}neisen parameters and $dN/dt$ is the rate of change of the phonon population, which depends on the electron temperature $T_e$ and phonon wavevector, k. The sum is over the entire Brillouin zone. Using the recently derived Eliashberg theory for laser-heated electron systems \cite{Carva}, we compute \textit{ab initio} $dN (T_e, \bm{k})/dt$  for Fe, which is shown  along $\Gamma -$H in Fig.\ \ref{fig:f3}(b).  Evidently, the phonon-population rate increases strongly with electron temperature and depends nonlinearly on the phonon wavevector. 


The depth-dependent strain profile of the forward moving strain wave can be approximated as  $\eta (z,t) =
\frac{1}{2 v^2 \rho} \int_{-\infty}^{+\infty} {\rm sgn}[z- v (t-t')] \frac{\partial \sigma}{\partial t} |_{\tiny{ (z',t')}} dt'$ , where $\rho$ is the density, $v$ the sound velocity, and $z'=| z -v(t-t')|$ \cite{Thomsen}. The \textit{ab initio} computed non-equilibrium strain profile is compared with that of the 2TM in Fig.\ \ref{fig:f3}(c). We computed an equilibrium electron-phonon coupling $G=$ $1.04 \times 10^{18}$\,Wm$^{-3}$K$^{-1}$ for the latter. For illustrative purposes we have excluded the electronic stress and assumed a uniform uniaxial excitation in a bulk material. The non-equilibrium strain profile (with $\gamma_e$=0) exhibits a sharply peaked front similar to that produced by the 2TM when including the electronic stress ($\gamma_e$=4.4, Fig.\ \ref{fig:f2}(b)). In contrast, the 2TM yields a flat, step-like strain profile for $\gamma_e$=0. These calculations underline the limitations of the 2TM, which appears to be insufficient to describe the impulsive excitation of acoustic phonons on picosecond time scales. A better knowledge of the energy transfer between electrons and lattice would be relevant not only for strain generation, but also for ultrafast demagnetization and other situations where the 2TM is employed \cite{Koopmans}. We note that time-resolved diffuse scattering can yield the evolution of non-equilibrium phonon populations \cite{Trigo2010}.

In conclusion, we measure time-resolved x-ray diffraction from a Fe/MgO film following demagnetization by femtosecond optical irradiation. We observe THz frequency oscillations in diffracted intensity in regions of reciprocal space corresponding to scattering from the individual coherent longitudinal acoustic phonons modes that make up the strain wave generated by the optical pulse. By fitting this data with a two-temperature thermo-elastic model including the thermal stress, we infer the spatial and temporal profiles of the strain. We find that the wavefront is $\sim$100 fs, comparable to electronic and spin thermalization timescales. This requires the total stress to increase much faster than the heating of the lattice. While the electron system in the two-temperature model can provide such a stress, we find that a nonthermal phonon system could also be responsible for the high-frequency Fourier components in the strain. First principles calculations suggest the phonon system is non-thermal on the timescale of our experiment and therefore has no temperature as assumed in the two-temperature model.


\begin{acknowledgements}
Research at SLAC was supported through the SIMES Institute which like the LCLS and SSRL user facilities is funded by the Office of Basic Energy Sciences of the U.S. Department of Energy under Contract No. DE-AC02-76SF00515. K.C., P.M., and P.M.O. acknowledge support from the European Community's Seventh Framework Program (FP7/2007-2013) under Grant Agreement No. 281043, FemtoSpin, by the Swedish Research Council (VR) and SNIC. T.H. acknowledges support by the LCLS. S.B. acknowledges support from the Knut and Alice Wallenberg Foundation. K.C. acknowledge support from the Czech Science Foundation (Grant No. 15-08740Y).
\end{acknowledgements}

\bibliography{ref.bib}
\end{document}


\section{Supplemental Material}
   
      \section*{Sample Preparation}
   The Fe layer was deposited on a MgO substrate that was cleaned with acetone and methanol using ultrasonic bath for 10 min in each solvent and then annealed at 500$^{\circ}$C for 1 hour in the deposition chamber at pressure $ < 10^{-9}$ torr. The MgO substrate was next cooled to 200$^{\circ}$C and exposed to atomic oxygen for 10 min. The 23 nm Fe film was deposited at substrate temperature of 50$^{\circ}$C from an e-beam source and then annealed at 350$^{\circ}$C for 1 hour. \emph{In situ} RHEED and \emph{ex-situ} x-ray reflectivity measurements indicated the Fe film was single crystalline, epitaxial, and smooth. The Fe layer was then capped with 3 nm MgO layer deposited by reactive magnetron sputtering to prevent its oxidation in air.
   
   \section*{Two-Temperature Model}
   
   We simulated the temperature profiles of the electronic and lattice subsystems by numerically solving ~\ref{eq:Te}and~\ref{eq:Tl} \cite{wright1994}.
\begin{equation}
C_e(T_e)\frac{\partial T_e}{\partial t} = 
\frac{\partial}{\partial z} (\kappa \frac{\partial T_e}{\partial z})
- G (T_e - T_l) + P(z,t) 
\label{eq:Te} \tag{s1}
\end{equation}

\begin{equation}
C_l \frac{\partial T_l}{\partial t} = G(T_e - T_l)
\label{eq:Tl} \tag{s2}
\end{equation}

where $P$ is the deposited laser power density, C is heat capacity, T is temperature, G is the electron-phonon coupling constant, and the subscripts e and l refer to the electronic and lattice subsystems. The resulting time-dependent temperatures were used to calculate the stress (given by eq bla bla bla in the main text) and thereby strain using a thermoelastic model described previously \cite{thomsen1986} \cite{wright1994}. Integration of the strain yields the displacements of the atomic planes, which are used in the next section to simulate the diffraction.

   \section*{Kinematic Diffraction Model}
   
   Consider the crystalline iron film in the conventional basis (simple cubic, 2 ions per unit cell). The basis atoms are labeled by the index $j$ and the unit cells are labeled by integers $n_x$, $n_y$, and $n_z$. We take all atomic motion to be along the sample normal, which we will call the z direction, and atomic planes are assumed to move in unison. Thus each atom's position in the z direction, $r_{z,j}^{n_z}$, depends upon $n_z$ and $j$, but not $n_x$ or $n_y$. The atomic positions, $\vec{r}_{j}(n_x, n_y, n_z)$ are given by 
\begin{equation}
\vec{r}_{1}(n_x, n_y, n_z) = a n_x \hat{x} + a n_y \hat{y} + r_{z}^{n_z,1}\hat{z}
\label{eq:r1} \tag{s3}
\end{equation}

\begin{equation}
\vec{r}_{2}(n_x, n_y, n_z) = a (n_x+\frac{1}{2}) \hat{x} + a (n_y+\frac{1}{2}) \hat{y} + r_{z}^{n_z,2}\hat{z}
\label{eq:r2} \tag{s4}
\end{equation}
 
where a is the lattice constant. In the limit of infinite x-ray penetration depth and no Debye-Waller effects,the complex electric field of the scattered x-rays is proportional to 

\begin{equation}
E \propto
\sum\limits_{j}
\sum\limits_{n_x,n_y,n_z}
exp[2 \pi i \vec{Q} \cdot \vec{r}_{j}(n_x, n_y, n_z)]
\label{eq:kdiff} \tag{s5}
\end{equation}

\begin{equation}
=
(\sum\limits_{n_x,n_y}
exp[2 \pi i a(n_x Q_x + n_y Q_y)])
\sum\limits_{n_z}^{N_z}(
exp[2 \pi i Q_z r_{z}^{n_z,1}] + 
exp[2 \pi i Q_z r_{z}^{n_z,2} + 
2 \pi i a (\frac{1}{2} Q_x + \frac{1}{2} Q_y)]
)
\label{eq:kdiffexpanded} \tag{s6}
\end{equation}

where $\vec{Q}$ is given by the difference in the wavevectors of the scattered and incident x-rays, $\vec{k_s}-\vec{k_i}$ where $\vec{k}$ is defined to have magnitude $\lambda^{-1}$ and $\lambda$ is the x-ray wavelength. The $\vec{Q}$ are determined for each detector pixel based on the experimental geometry.

Upon taking the modulus square of the above and the limit that the atomic planes are infinite in the x and y directions, the sum over $n_x$ and $n_y$ yields sharp peaks which are periodic in $Q_x, Q_y$ \cite{warren}. In our experiment, the width of these peaks is determined by the bandwidth of the LCLS pink beam. As such, we replace

\begin{equation}
=
|(\sum\limits_{n_x,n_y}
exp[2 \pi i a(n_x Q_x + n_y Q_y)])|^2 \rightarrow
exp[-\frac{q_{xy}^2}{\alpha^2}]
\label{eq:xywidth} \tag{s7}
\end{equation}

where $q_{xy}$ is the magnitude of the reduced scattering vector in the xy plane and $\alpha$ is the LCLS wavevector bandwidth, $\approx \Delta \lambda/\lambda^2$.
Utilizing the above substitution and inserting the Debye-Waller factor and finite x-ray penetration depth,~\ref{eq:kdiffexpanded} reduces to
   
\begin{equation}
E \propto
e^{-\frac{q_{xy}^2}{2 \alpha^2}} 
\sum\limits_{j=1,2} 
\sum\limits_{n_z=1}^{N_z} 
exp[ 
-B(T_l^{n_z,j} ) s^2 + 
(i 2 \pi Q_z  - \frac{1}{d_x} )
r_{z}^{n_z,j}+ i \phi_{Q} \delta_{j,2}
]
\label{eq:finalE} \tag{s8}
\end{equation}

where $\phi_Q$ is defined as
\begin{equation}
\phi_Q \equiv 2 \pi a (\frac{1}{2} Q_x + \frac{1}{2} Q_y)
\label{eq:defphiq} \tag{s9}
\end{equation}

$T_i^{n,j}$ is the lattice temperature for basis atom j of atomic plane n,  and $d_x$ is the x-ray penetration depth. $exp[-B(T_i^j)s^2]$ is the Debye-Waller factor where $B$ is the temperature factor, s is $Sin(\theta)/\lambda$ and $\theta$ is the scattering angle.



\begin{figure}[H]
\centering
\includegraphics[width=9cm]{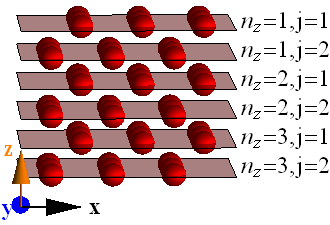}
\caption{Illustration of atomic plane labeling scheme used for calculating diffraction patterns from the BCC iron film. $z$ is the sample normal. }
\end{figure}

   \section*{Model Variables and Parameters}

\noindent
\begin{table}[H]
   \begin{center}

      \begin{tabular}{| c | p{7cm} | c | c | c |}
      \hline
      
      Parameter & Description & Value & Reference \\ \hline
      
      $a$ & lattice constant for BCC iron & 2.87 \AA  & \\ \hline
      
      $\alpha$ & 
	Wave-vector bandwidth of the LCLS pink beam, given
	approximately by $\frac{\Delta \lambda}{\lambda^2}$ where $\Delta \lambda$ is the wavelength bandwidth. We estimated this based on a FWHM bandwidth of 0.5\% & 
	0.0025 $\textup{\AA}^{-1}$ & ~\ref{eq:finalE} \\ \hline
    
    $B(T)$ & Debye-Waller temperature factor & & \cite{DWfactors} \\ \hline
    
    $s$ & $s = Sin(\theta)/\lambda$ where $\theta$ is the scattering angle and $\lambda$ is the x-ray wavelength & & 
    ~\ref{eq:finalE} \\ \hline
    
    $E$ & Complex electric field of scattered X-rays & & 
    ~\ref{eq:finalE} \\ \hline
    
    $\phi_Q$ & $\phi_Q = 2 \pi (Q_x + Q_y)a/2$ is the extra phase accumulated by x-rays scattering off the second basis atom. & & 
    ~\ref{eq:finalE} \\ \hline
    
    $r_z^{n_z,j}$ & Sample-normal displacement of atomic plane $n_z$,$j$. 
    & & ~\ref{eq:finalE} \\ \hline
    
    $T_l^{n_z,j}$ & Lattice temperature of atomic plane $n_z$,$j$.
    & & \ref{eq:finalE} \\ \hline

    G & Electron-Phonon Coupling Constant & 5.48e18 $W m^{-3} K^{-1}$
    & \cite{lin2008} \cite{e-pconstweb2} \\ \hline
    
    $C_e$ & Electronic heat Capacity per unit temperature & 
    702 $J m^{-3} K^{-2}$ & \cite{kittel} \\ \hline
    
    $C_l$ & Lattice heat capacity & 3.33e6 $J m^{-3} K^{-1}$ & 
    \cite{latticeC} \\ \hline
    
    $\kappa$ & Electronic thermal conductivity & 80 $W m^{-1} K^{-1}$
    & \cite{kapparef} \\ \hline
    
    $K$ & Iron bulk Modulus & 1.7e11 $N m^{-2}$ & \cite{bulkmodulus}
    \\ \hline
    
    $\beta$ & Iron linear thermal expansion coefficient & 
    1.2e-5 $K^{-1}$ & \cite{ashcroftmermin} \\ \hline
    
    $\gamma_l$ & 
    Iron lattice Gr\"{u}neisen parameter, given by $3 K \beta / C_l$ 
    & 1.8 & \\ \hline
    
    $\rho_{Fe}$ & Iron density & 7.87e3 $kg m^{-2} s^{-1}$ & 
    \cite{FeDensity} \\ \hline
    
    $c_{Fe}$ & Iron speed of sound & 5.130e3 $m s^{-1}$ & 
    \cite{crc} \\ \hline
    
    $Z_Fe$ & 
    Iron acoustic impedance $ =\rho_{Fe} c_{Fe}$ & 40.4 $kg m^{-2}s^{-1}$ 
    & \\ \hline
    
    $d_o$ & optical penetration depth & 17.5 nm & 
    \cite{optPenDepth} \\ \hline
    
    $\rho_{MgO}$ & MgO density & 3.58e3 $kg m^{-2} s^{-1}$ & 
    \cite{mgoImp} \\ \hline
    
    $c_{MgO}$ & MgO speed of sound & 9.1e3 $m s^{-1}$ &
    \cite{mgoImp} \\ \hline
    
    $Z_{MgO}$ & 
    MgO acoustic impedance $ =\rho_{MgO} c_{MgO}$ & 32.6 $kg m^{-2}s^{-1}$ 
    & \\ \hline

\end{tabular}

\end{center}
\caption{Description of variables and constants used in simulations of diffraction and strain and temperature profiles.}
\end{table}

\noindent
\begin{table}[H]
   \begin{center}

      \begin{tabular}{| c | p{7cm} | c | c | c |}
      \hline
      
      Parameter & Description & Best Fit for $\gamma_e=0$ & 
      Best Fit $\gamma_e$ free\\ \hline
      
      $\gamma_e$ & Electronic Gr\"{u}neisen parameter & 
      Fixed at 0 & 4.4 \\ \hline
      
      $N_z$ & Film thickness in unit cells (Film thickness/a) & 
      79 & 79 \\ \hline
      
      F & Absorbed Fluence & 0.85 $mJ cm^{-2}$ & 
      0.66 $mJ cm^{-2}$ \\ \hline

      Grazing Angle ($d_x$) & X-ray grazing angle, which determines the x-ray penetration depth ($d_x$) & 
      0.435 deg (54 nm) & 0.430 deg (46 nm) \\ \hline
      
       & Offset in sample azimuthal angle & -0.58 deg  & 
       -0.57 deg \\ \hline
       
       & offset in arrival time of optical pulse & -48 fs & 
       -7 fs \\ \hline
       
       red$\chi^2$ & Reduced chi squared & 
       12.2 & 9.1 \\ \hline

\end{tabular} 

\end{center}
\caption{Free parameters used in fit when including and excluding stress from electronic subsystem.}
\end{table}


\bibliography{ref.bib}